\documentclass[journal]{IEEEtran}
\ifCLASSINFOpdf
\else
\fi

\usepackage{epsfig}
\usepackage{graphicx,color,psfrag,subfigure}
\usepackage{graphicx}
\usepackage{color}
\usepackage{placeins}
\usepackage{float}
\usepackage{tabularx,colortbl}
\usepackage{amsmath} 
\usepackage[table]{xcolor}
\usepackage{hhline}

\begin{document}
\title{Gaussian Process Regression for Arctic Coastal Erosion Forecasting}

\author{Matthew~Kupilik,~\IEEEmembership{Member,~IEEE,}
        Frank~Witmer, Euan-Angus~MacLeod, Caixia Wang, Tom Ravens
        % <-this % stops a space
\thanks{Matthew Kupilik is with the Department
of Electrical Engineering, University of Alaska Anchorage, 3211 Providence Drive, Anchorage AK, 99508 USA e-mail: mkupilik@alaska.edu}% <-this % stops a space
\thanks{Frank Witmer is with the Computer Science and Computer Engineering Department at the University of Alaska Anchorage}% <-this % stops a space
\thanks{Euan-Angus MacLeod is with the Civil Engineering Department at the University of Alaska Anchorage}% <-this % stops a space
\thanks{Caixia Wang is with the Geomatics Department at the University of Alaska Anchorage}%
\thanks{Tom Ravens is with the Civil Engineering Department at the University of Alaska Anchorage}%

}


\markboth{submitted to the IEEE Transactions on Geoscience and Remote Sensing}%
{Shell \MakeLowercase{\textit{et al.}}: Bare Demo of IEEEtran.cls for IEEE Journals}



\maketitle

\begin{abstract}
Arctic coastal morphology is governed by multiple factors, many of which are affected by climatological changes. As the season length for shorefast ice decreases and temperatures warm permafrost soils, coastlines are more susceptible to erosion from storm waves. Such coastal erosion is a concern, since the majority of the population centers and infrastructure in the Arctic are located near the coasts.  Stakeholders and decision makers increasingly need models capable of scenario-based predictions to assess and mitigate the effects of coastal morphology on infrastructure and land use.  Our research uses Gaussian process models to forecast Arctic coastal erosion along the Beaufort Sea near Drew Point, AK.  Gaussian process regression is a data-driven modeling methodology capable of extracting patterns and trends from data-sparse environments such as remote Arctic coastlines.  To train our model, we use annual coastline positions and near-shore summer temperature averages from existing datasets and extend these data by extracting additional coastlines from satellite imagery.  We combine our calibrated models with future climate models to generate a range of plausible future erosion scenarios. Our results show that the Gaussian process methodology substantially improves yearly predictions compared to linear and nonlinear least squares methods, and is capable of generating detailed forecasts suitable for use by decision makers. 
\end{abstract}


% \fi
\IEEEpeerreviewmaketitle

\section{Introduction}
\IEEEPARstart{A}{rctic} coastlines are experiencing high rates of erosion.  The loss of Arctic coastal land has a significant effect both on the large proportion of the Arctic population that resides along the coast as well as the military and energy production infrastructure.  Several Arctic Alaska communities require relocation due to the almost complete loss of land, several are facing threats to sanitation and transportation infrastructure.
%[FW: the prior sentence needs a citation else should be cut or toned down]
To meet these challenges in a cost effective manner, community and industry planners require erosion forecasting on an annual to decadal temporal scale and a spatial scale sufficient to plan for the built environment, such as roads, buildings and airstrips.  Making accurate forecasts is further complicated by climate change effects, and uncertainties about what temperatures will be in future years.  

Approaches to modeling coastal morphology vary widely in scale, both temporally and spatially.  Models used for long-term infrastructure planning and climate impacts at the country level typically require annual or decadal temporal scales and spatial scales of $1$ to $100$ kilometers.  In contrast, models used for managing local infrastructure need hourly or daily temporal scales and spatial scales on the order of a single meter.  These widely different scales and the high variability in coastal dynamics have resulted in a wide range of modeling approaches.  Models for coastal forecasting can be categorized based on the spatial and temporal scales they attempt to forecast over.

High-order physics-based models involving large numbers of inputs have been created for specific sets of coastal geologies \cite{roelvink2010xbeach,Jamal2014}.  In order to provide a reasonable spatial resolution such models require detailed knowledge of grain size, hydrodynamics, water surface elevation data, bluff and beach profile and offshore bathymetry.  Most of these inputs are difficult to both measure and calculate using atmospheric or meteorological modeling.  Model complexity tends to limit forecasting using such models to small sections of coastline and to specific erosion inducing events (such as flooding or storms).  To obtain reasonable accuracy such models also require expert knowledge in estimating inputs and knowledge over what coasts are suitable for modeling \cite{Thieler2000}.  Complex process based models are most useful for modeling small sections of coastline at a very fine spatial and temporal scale.  

In order to meet the need for annual to decadal forecasting over tens of kilometers or even larger sections of coastline, empirical or combined empirical and process based approaches are used. Such data-driven models take advantage of the growing availability of good-quality coastline data. Recent work has attempted to integrate such models of disparate scope and methodology, but widely applicable multi-scale coastal erosion prediction remains a difficult problem \cite{Maanen2016}.

There is a growing body of Arctic coastal erosion research that uses process models \cite{Barnhart2014,Ravens2012,Ping2011,Lantuit2012}, though these are less common in the Arctic than more populated areas \cite{Maanen2016}.  The main limiting factor for developing data-rich process models is the dearth of detailed data inputs for Arctic study sites. Regions with well-developed multi-decadal coastal position data have seen significant gains made in prediction using data approaches \cite{Burningham2017} or empirical models combined with process modeling \cite{Corbella2012, VitousekS.BarnardP.L.LimberP.EriksonL.&Cole2017}.

Applications in the Arctic face additional modeling challenges.  For process models, dynamics due to reduced sea ice and increased temperatures are poorly understood.  For data-driven models, the Arctic has fewer observations, less comprehensive satellite coverage, and higher costs for obtaining high precision coastal measurements in remote regions.  These challenges mean the regions facing the most immediate threats from coastal erosion are also least able to predict them \cite{Jones2008}.  

Our work applies a data-driven modeling approach using Gaussian process regression. We estimate multiple models using observations of past coastlines and hindcasted environmental data and evaluate their relative predictive capabilities.  In particular, we compare Gaussian process models with linear and nonlinear regression models.  The formulation of these models means that we are comparing a linear empirical modeling technique to a Bayesian empirical modeling technique.  This comparison allows us to determine if there is sufficient information in the coastline data set to make annual predictions using modern estimation methods.

From our baseline models calibrated from the historical observed data, we then forecast future coastlines 5 and 10 years into the future.  Our forecasts rely on a future scenarios modeling approach \cite{shearer2005,Mahmoud2009,witmer2017} that incorporates temperature data from multiple future climate models. This allows us to generate multiple plausible future coastlines that are sensitive to which future climate conditions actually materialize.

%Comparison of the Gaussian process methodology to process-based models is a future goal of the project, as is testing the predictive power of additional covariates such as sea ice extent, aggregated wave and storm information, and air and water temperature [this last sentence seems more suited for the conclusion].

%We build on existing work from both empirical and process-based modeling [how are we building on existing process-based work? perhaps this paragraph should be moved to the beginning of the intro and expanded to explain the 2 approaches].

\section{Model Data}
The accuracy of any machine learning or empirical technique is closely connected to the quality of data used to train it.  Dynamics that have not been observed will not be present in the predictive model.  For larger data sets, determining a model structure that can capture all the patterns present in the data is often the most challenging aspect. In this section, we describe our coastline data, the generalization process we use to measure coastline change, and the additional environmental data we use as both model covariates and to generate future scenarios.

\subsection{Coastline Position Data}
\begin{figure}[h]
\begin{center}
\noindent
  \resizebox{8cm}{!}{\includegraphics{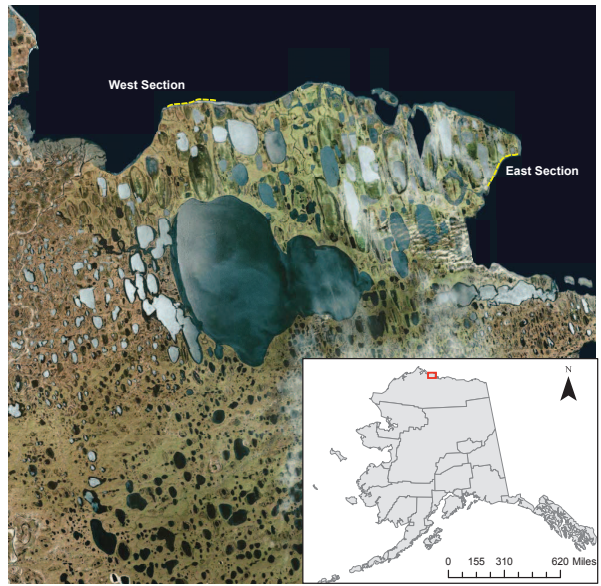}}
	\caption{Study area and analyzed subsections.}\label{fig:drewmap}
\end{center}
\end{figure}

Our Arctic study area spans the coastline from approximately Drew Point to Cape Halkett, Alaska (Figure \ref{fig:drewmap}).
Our sources for coastal position data included aerial and satellite imagery from several public datasets as well as GPS coastline surveys conducted by other researchers; the source, resolution, and year for each is shown in Table \ref{tab:Table1Label}.  Years with GPS measurements were produced by \cite{Barnhart2014} and have a stated accuracy of $1-2$ meters.  Coastline segments for 1947 and 2002 were generated via satellite and aerial imagery \cite{Gibbs2015}.  Years 1955, 1979, and 2007 were generated from aerial photography by \cite{Jones2009,Jones2008}.  To these existing data, we digitized coastlines for two more years, 2009 and 2016, by manually photointerpreting SPOT 5 and Landsat 8 satellite imagery. All the measurements were collected during the summer months using visible bluff lines. Due to the varying data sources, not all transect locations have measurements at all years.
\begin{table}[h]
\begin{center}
\caption{Coastal Position Data} \label{tab:Table1Label}
\begin{tabular}{lcc}

\cellcolor[gray]{0.8}Year & \cellcolor[gray]{0.8}Source & \cellcolor[gray]{0.8}Resolution \\ 
 \hline
 1947 & NOAA Topographic Sheets & 15.7 m   \\
 1955 & Aerial Photography & 2.5 m \\
 1979 & Aerial Photography & 2.5 m \\
 2002 & USGS Orthophoto Quads & 6 m  \\
 2007 & Aerial Photography & 2.5 m \\
 2008 & GPS & 1-2 m  \\
2009 & Spot 5 & 5 m  \\
2011 & GPS & 1-2 m  \\
2012 & GPS & 1-2 m  \\
2016 & LandSat 8 & 15 m  \\
 \hline
\end{tabular}
\end{center}
\end{table} 

\begin{figure}[!t]
\centering
%\noindent
  \resizebox{8cm}{!}{% This file is generated by the MATLAB m-file laprint.m. It can be included
% into LaTeX documents using the packages graphicx, color and psfrag.
% It is accompanied by a postscript file. A sample LaTeX file is:
%    \documentclass{article}\usepackage{graphicx,color,psfrag}
%    \begin{document}\input{yearHist}\end{document}
% See http://www.mathworks.de/matlabcentral/fileexchange/loadFile.do?objectId=4638
% for recent versions of laprint.m.
%
% created by:           LaPrint version 3.16 (13.9.2004)
% created on:           12-Oct-2017 18:41:20
% eps bounding box:     15 cm x 12.0787 cm
% comment:              
%
\begin{psfrags}%
\psfragscanon%
%
% text strings:
\psfrag{s02}[t][t]{\color[rgb]{0.15,0.15,0.15}\setlength{\tabcolsep}{0pt}\begin{tabular}{c}Number of Measurement Years\end{tabular}}%
\psfrag{s03}[b][b]{\color[rgb]{0.15,0.15,0.15}\setlength{\tabcolsep}{0pt}\begin{tabular}{c}Number of Transects\end{tabular}}%
%
% axes ticklabel color:
\color[rgb]{0.15,0.15,0.15}%
%
% xticklabels:
\psfrag{x01}[t][t]{3}%
\psfrag{x02}[t][t]{4}%
\psfrag{x03}[t][t]{5}%
\psfrag{x04}[t][t]{6}%
\psfrag{x05}[t][t]{7}%
\psfrag{x06}[t][t]{8}%
\psfrag{x07}[t][t]{9}%
\psfrag{x08}[t][t]{10}%
%
% yticklabels:
\psfrag{v01}[r][r]{0}%
\psfrag{v02}[r][r]{20}%
\psfrag{v03}[r][r]{40}%
\psfrag{v04}[r][r]{60}%
\psfrag{v05}[r][r]{80}%
\psfrag{v06}[r][r]{100}%
\psfrag{v07}[r][r]{120}%
\psfrag{v08}[r][r]{140}%
\psfrag{v09}[r][r]{160}%
\psfrag{v10}[r][r]{180}%
%
% Figure:
\resizebox{12cm}{!}{\includegraphics{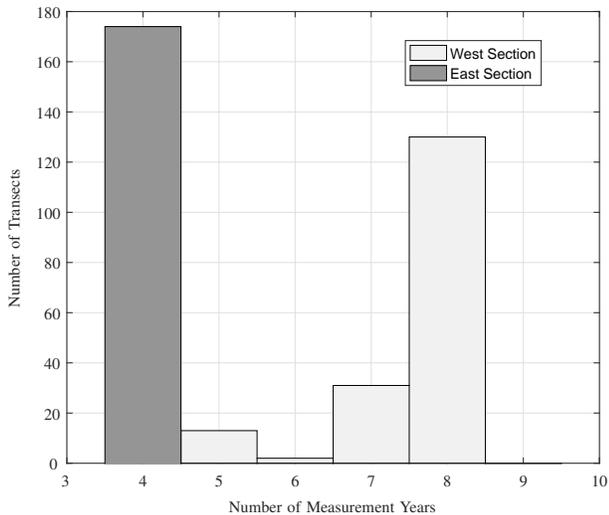}}%
\end{psfrags}%
%
% End yearHist.tex
}
\caption{Histogram showing number of years available for each section.}
\label{fig:yearhist}
\end{figure}

Restricting our analysis to sections of coastlines that are outlier free and have sufficient measurements (at least four years) results in two sections, both shown in Figure \ref{fig:drewmap}. The western section closest to Drew Point is approximately 9 km long and has the highest concentration of data as it has been the study site for several Arctic erosion projects.  For this section, the number of years of measurement data varies from 5 to 8, with most having 8 years available (Figure \ref{fig:yearhist}). The baseline year of 1947 does not count in this total since it is the reference year, and the last year, 2016, is excluded since it is used for prediction.  The eastern section is also approximately 9 km long and has fewer position measurements available (Figure \ref{fig:yearhist}), with four measurements available per transect (1955, 1979, 2002, 2007).  The number of annual coastline measurements for our study area is much fewer than other works that applied data based estimation to coastal forecasting models \cite{VitousekS.BarnardP.L.LimberP.EriksonL.&Cole2017}.

\subsection{Measuring Coastline Change}
In order to reduce the dimensionality of the estimation problem, we use a modeling method similar to the standard one-line approaches often used for coastal erosion. We achieve this by discretizing our coastline using a set of transects chosen to be orthogonal to the coastline direction at the point.  We make use of the $50$ meter transects generated in \cite{Gibbs2015} and extend them where necessary to intersect our newly added coastline data. We establish a baseline coastal position from the earliest data source (1947) as the intersection of each baseline to the coastline for that year. Change in coastline is measured by the distance from that baseline point, with negative distances indicating erosion and a positive distances indicate deposition. Distance calculations are in meters since all coastline and transect data are projected to UTM zone 5N.

In order to capture effects {\it along} the coastline such as longshore drift, we explicitly build in a spatial dependence parameter to the model. This parameter is calculated by assigning the west-most baseline point coastal position zero and recursively adding the distance between each neighboring point as one travels east along the coast. This produces a continuous distance measure between baselines for non-continuous coastline segments that allows us to capture the spatial dependence between transects.%The input is then two dimensional, consisting of the year and the distance along the baseline coast. [I don't know what this last sentence means -- delete it?]

\begin{figure*}[!t]
\centering
\noindent
  \resizebox{16cm}{!}{% This file is generated by the MATLAB m-file laprint.m. It can be included
% into LaTeX documents using the packages graphicx, color and psfrag.
% It is accompanied by a postscript file. A sample LaTeX file is:
%    \documentclass{article}\usepackage{graphicx,color,psfrag}
%    \begin{document}\input{EastWest3D}\end{document}
% See http://www.mathworks.de/matlabcentral/fileexchange/loadFile.do?objectId=4638
% for recent versions of laprint.m.
%
% created by:           LaPrint version 3.16 (13.9.2004)
% created on:           12-Oct-2017 20:57:32
% eps bounding box:     20 cm x 9.5461 cm
% comment:              
%
\begin{psfrags}%
\psfragscanon%
%
% text strings:
\psfrag{s01}[lt][lt]{\color[rgb]{0.15,0.15,0.15}\setlength{\tabcolsep}{0pt}\begin{tabular}{l}Position \\ (meters $\times 10^{5}$)\end{tabular}}%
\psfrag{s02}[b][b]{\color[rgb]{0.15,0.15,0.15}\setlength{\tabcolsep}{0pt}\begin{tabular}{c}Distance from 1947 Baseline (meters)\end{tabular}}%
\psfrag{s03}[b][b]{\color[rgb]{0.15,0.15,0.15}\setlength{\tabcolsep}{0pt}\begin{tabular}{c}Distance from 1947 Baseline (meters)\end{tabular}}%
\psfrag{s04}[lt][lt]{\color[rgb]{0.15,0.15,0.15}\setlength{\tabcolsep}{0pt}\begin{tabular}{l}Year\end{tabular}}%
\psfrag{s05}[b][b]{\color[rgb]{0,0,0}\setlength{\tabcolsep}{0pt}\begin{tabular}{c}East Section\end{tabular}}%
\psfrag{s06}[rt][rt]{\color[rgb]{0.15,0.15,0.15}\setlength{\tabcolsep}{0pt}\begin{tabular}{r}Year\end{tabular}}%
\psfrag{s07}[t][t]{\color[rgb]{0.15,0.15,0.15}\setlength{\tabcolsep}{0pt}\begin{tabular}{c}Position \\ (meters $\times 10^{5}$)\end{tabular}}%
\psfrag{s08}[b][b]{\color[rgb]{0,0,0}\setlength{\tabcolsep}{0pt}\begin{tabular}{c}West Section\end{tabular}}%
%
% axes ticklabel color:
\color[rgb]{0.15,0.15,0.15}%
%
% xticklabels:
\psfrag{x01}[t][t]{3}%
\psfrag{x02}[t][t]{3.1}%
\psfrag{x03}[t][t]{}%
\psfrag{x04}[t][t]{2.1}%
\psfrag{x05}[t][t]{2.2}%
\psfrag{x06}[t][t]{}%
%
% yticklabels:
\psfrag{v01}[r][r]{1950}%
\psfrag{v02}[r][r]{2000}%
\psfrag{v03}[r][r]{2050}%
\psfrag{v04}[r][r]{1950}%
\psfrag{v05}[r][r]{2000}%
\psfrag{v06}[r][r]{2050}%
%
% zticklabels:
\psfrag{z01}[r][r]{-500}%
\psfrag{z02}[r][r]{-400}%
\psfrag{z03}[r][r]{-300}%
\psfrag{z04}[r][r]{-200}%
\psfrag{z05}[r][r]{-100}%
\psfrag{z06}[r][r]{0}%
\psfrag{z07}[r][r]{100}%
\psfrag{z08}[r][r]{200}%
\psfrag{z09}[r][r]{-1200}%
\psfrag{z10}[r][r]{-1000}%
\psfrag{z11}[r][r]{-800}%
\psfrag{z12}[r][r]{-600}%
\psfrag{z13}[r][r]{-400}%
\psfrag{z14}[r][r]{-200}%
\psfrag{z15}[r][r]{0}%
%
% Figure:
\resizebox{16cm}{!}{\includegraphics{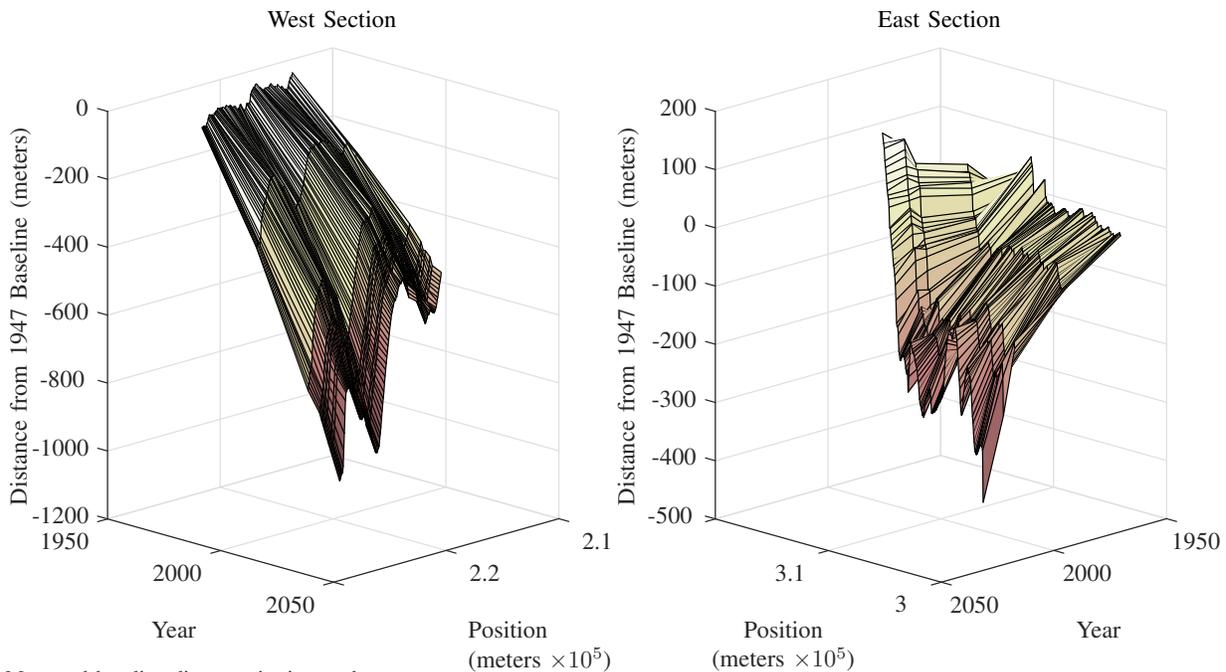}}%
\end{psfrags}%
%
% End EastWest3D.tex
}
  \caption{Measured baseline distances in time and space.}
	\label{fig:EastWest3D}
\end{figure*}

%This dimension reduction approach limits our approach to modeling approximately 10 km sections of coastline individually [why?  I don't understand this previous statement and it seems unrelated to the rest of this paragraph -- so delete?].
Since the coastline is being discretized into 50 m segments, as the coastline curves the orthogonal transects overlap.  When future coastal measurements are assigned to transects they can oscillate between erosion and deposition, resulting in outliers.  We attempt to minimize this effect by spatially smoothing using averages.  With so few coastline measurements the presence of any outliers results in large increases in error for model forecasts. In order to fairly capture the predictive ability of our Gaussian process approach, we exclude these problematic sections of coastline from our model. % [begin delete...] resulting in the two sample coastline sections shown in Figure \ref{fig:drewmap} [...end delete?]. 
Figure \ref{fig:EastWest3D} shows the discretized coastal distance measurements from the 1947 baseline plotted by transect distance along the coast for each year of observation.  The west section shows continually increasing rates of erosion, with no deposition at any transects.  The east section shows deposition at some transects and a much larger spatial variation in erosion rates.

\subsection{Environmental Data}
To improve our model accuracy, we incorporate relevant available environmental data. Factors such as near-shore water temperature, wind, wave height, and wave direction are strong predictors of Arctic erosion \cite{Ravens2012, Barnhart2014}.  However, directly measured environmental data for Arctic coastlines does not exist at the spatial resolutions we require. Instead, we use hindcasted model data of near-shore temperature measurements for learning and corresponding forecasted temperature data to generate our future scenarios.  The spatial and temporal resolution of other relevant environmental variables is a significant hurdle, and the only measurement we use is near-shore water temperature, calculated for all model years up to 2016 using model generated data.
%[I think we should replace the VanVuuren2007 citation with this one from 2011:]
%Van Vuuren, Detlef; Jae Edmonds, Mikiko Kainuma, Keywan Riahi, Allison Thomson, Kathy Hibbard, George C Hurtt, Tom Kram, Volker Krey, Jean‐Francois Lamarque, Toshihiko Masui, Malte Meinshausen, 31 Nebojsa Nakicenovic, Steven J Smith & Steven K Rose (2011) The representative concentration pathways: An overview. Climatic Change 109(1): 5‐31.

\begin{figure}[!t]
\centering
%\noindent
  \resizebox{8cm}{!}{% This file is generated by the MATLAB m-file laprint.m. It can be included
% into LaTeX documents using the packages graphicx, color and psfrag.
% It is accompanied by a postscript file. A sample LaTeX file is:
%    \documentclass{article}\usepackage{graphicx,color,psfrag}
%    \begin{document}\input{hindTemp}\end{document}
% See http://www.mathworks.de/matlabcentral/fileexchange/loadFile.do?objectId=4638
% for recent versions of laprint.m.
%
% created by:           LaPrint version 3.16 (13.9.2004)
% created on:           12-Oct-2017 20:01:08
% eps bounding box:     15 cm x 10.407 cm
% comment:              
%
\begin{psfrags}%
\psfragscanon%
%
% text strings:
\psfrag{s03}[b][b]{\color[rgb]{0.15,0.15,0.15}\setlength{\tabcolsep}{0pt}\begin{tabular}{c}UTM Y (meters)\end{tabular}}%
\psfrag{s04}[t][t]{\color[rgb]{0.15,0.15,0.15}\setlength{\tabcolsep}{0pt}\begin{tabular}{c}UTM X (meters)\end{tabular}}%
%
% axes ticklabel color:
\color[rgb]{0.15,0.15,0.15}%
%
% xticklabels:
\psfrag{x01}[t][t]{4.6}%
\psfrag{x02}[t][t]{4.7}%
\psfrag{x03}[t][t]{4.8}%
\psfrag{x04}[t][t]{4.9}%
\psfrag{x05}[t][t]{5}%
\psfrag{x06}[t][t]{5.1}%
\psfrag{x07}[t][t]{5.2}%
\psfrag{x08}[t][t]{5.3}%
\psfrag{x09}[t][t]{\shortstack{5.4\\$\times 10^{5}\ $}}%
%
% yticklabels:
\psfrag{v01}[r][r]{7.82}%
\psfrag{v02}[r][r]{7.83}%
\psfrag{v03}[r][r]{7.84}%
\psfrag{v04}[r][r]{7.85}%
\psfrag{v05}[r][r]{7.86}%
\psfrag{v06}[r][r]{7.87}%
\psfrag{v07}[r][r]{7.88}%
\psfrag{ypower1}[Bl][Bl]{$\times 10^{6}$}%
%
% Figure:
\resizebox{12cm}{!}{\includegraphics{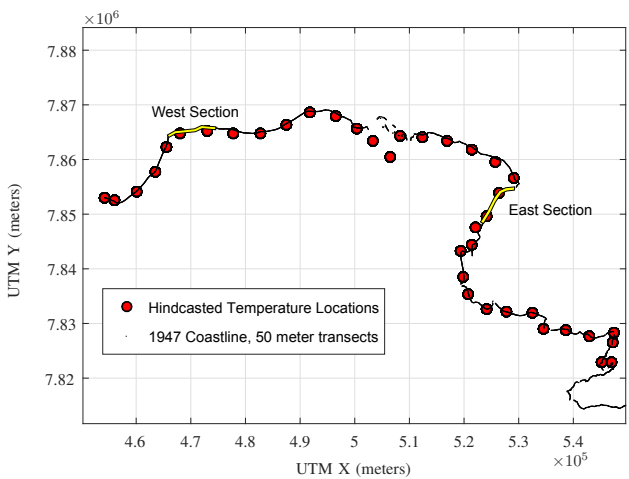}}%
\end{psfrags}%
%
% End hindTemp.tex
}
\caption{Spatial location of hindcasted near shore temperatures.}
\label{fig:hindtemp}
\end{figure}

In order to create plausible future temperature data, we apply the delta method \cite{Ramirez-Villegas2010} to five different global climate models shown in Table \ref{tab:forecastTempTable}.  We analyze two cases to demonstrate the ability of the Gaussian process method to forecast coastlines using near-shore temperature data.  The first uses a Representative Concentration Pathway (RCP) of 2.6 $\mbox{Wm}^{-2}$ \cite{vanVuuren2011} which represents our optimistic forecast scenario, an emissions peak and decline.  The second scenario we consider uses an RCP of 8.5 $\mbox{Wm}^{-2}$ \cite{Riahi2007} which represents a more pessimistic future scenario with higher levels of atmospheric CO$_2$ and increasing temperatures in the Arctic.  For both of these future scenarios, we assemble data from the five models and downsample the temperature to our observation points shown in Figure \ref{fig:hindtemp}.  We use linear interpolation to calculate the near-shore water temperature at each month for all coastal transects.  Since the temporal scale of our model is annual, the yearly average is taken as the mean of August and September, the months during which most erosion occurs.
\begin{table}[h]
\begin{center}
\caption{Forecast CMIP5/AR5 Model Sources} \label{tab:forecastTempTable}
\begin{tabular}{p{3.5cm} p{3cm}}
\cellcolor[gray]{0.8}Center  & \cellcolor[gray]{0.8}Model \\ 
 \hline
National Center for Atmospheric Research & CCSM4 Community Earth System Model 4 \cite{gent2011community} \\
\hline
NOAA Geophysical Fluid Dynamics Laboratory & GFDL-CM3 Coupled Model 3.0 \cite{Program2011} \\
 \hline
NASA Goddard Institute for Space Studies & GISS-E2-R ModelE/Russell \cite{schmidt2006present} \\
 \hline
Institut Pierre-Simon Laplace & IPSL Coupled Model v5A IPSL-CM5A-LR \cite{dufresne2013climate} \\
 \hline
Meteorological Research Institute & MRI-CGCM3 Coupled General Circulation Model v3.0 \cite{yukimoto2001new} \\
 \hline
\end{tabular}
\end{center}
\end{table}

\section{Methodology}
We use Gaussian process regression to model coastal erosion and compare its performance to other commonly applied erosion models. The dynamics that are captured during the Gaussian regression reflect the dynamics that have been observed in the past.  As time scales for prediction increase (e.g. decades or longer), major shifts in process types can result in poor prediction using purely data-driven approaches \cite{French2016}.  For instance, the erosion of a beach bluff may proceed at a predictable rate until it reaches a rocky cliff; or the erosion along a narrow spit may change drastically after the spit is breached.  These state changes have large impacts and are often either non-cyclical or repeat with such low frequency that predicting them is very challenging.  For the time scales and data observations in our study area, we observe few such major state changes.  There are several adjustments from beach and spit to coastal bluff but for most of our data set, we are predicting erosion for coastal bluffs exposed to open ocean and sea ice.  Other methods such as process-based models suffer from the same difficulty predicting these kinds of abrupt changes in system dynamics. 

\subsection{Gaussian Processes}
Gaussian processes provide a non-parametric model that can tractably be fit to existing data.  They have been extensively applied in the geostatistics community and are often referred to as Kriging.  Gaussian processes have rapidly gained traction in the machine learning community and have been applied in a variety of contexts \cite{CarlEdward2006}.  Their ability to describe relationships between the input and output variables without requiring a rigid model form makes them applicable to many problem classes.  We provide an brief introduction to Gaussian processes here; for a more thorough discussion, see \cite{CarlEdward2006}.

Instead of specifying the class of functions to fit, Gaussian processes define a distribution over the space of possible functions.  This distribution is defined as a collection of random variables, any finite number of which have a joint Gaussian distribution.  The Gaussian process is then completely defined by the form of a mean and covariance:
\begin{eqnarray}
m(\mathbf{x}) &=& \mathbf{E}[f(\mathbf{x})] \\
k(\mathbf{x},\mathbf{x}') &=& \mathbf{E}[(f(\mathbf{x})-m(\mathbf{x}))(f(\mathbf{x}')-m(\mathbf{x}'))] 
\end{eqnarray}
where $\mathbf{E}$ is the expectation operator.  The input vector is $\mathbf{x} \in \Re^D$ and the unknown function is $f$. %[what is x'?]
To fit the data we must make some assumptions on the form of the covariance, also known as a kernel. There are many choices of covariance functions and here we will describe them as a function parameterized by a vector $\mathbf{\theta}$.  If a valid covariance function is chosen, the resulting function $f(\mathbf{x}) \sim \mathcal{N}(\mathbf{\mu},K)$ is Gaussian with mean $\mu_i=m(x_i)$ and covariance $K_{ij} = k(x_i,x_j|\mathbf{\theta})$ for the $i$th and $j$th element of $\mathbf{x}$ .  The standard notation for combining the mean and covariance functions as a Gaussian process (GP) is:
\begin{equation}
f(\mathbf{x}) \sim \mathcal{GP}(m(\mathbf{x}),k(\mathbf{x},\mathbf{x}')).
\end{equation}    

The primary goal is to estimate the function $f$ from as many known instances of input ($\mathbf{x}$) and noisy observed output pairs ($\mathbf{y}+\epsilon$) that are available, and then make use of this estimate by predicting the value of the function ($f_*$) at other input values ($x_*$).  Given a dataset $\mathcal{D}=(\mathbf{y},\mathbf{x})$, we would like to predict the output at a new input $x_*$.  If we assign a prior and a likelihood, then we can condition the prior on our observations and calculate a posterior:
\begin{equation}
p(f|\mathbf{x},\mathbf{y}) \propto p(\mathbf{y}|\mathbf{x},f)p(f)
\end{equation}
This operation is mathematically equivalent to drawing random functions from the prior, and rejecting those that do not agree with the data.  If we assume the prior and likelihood are Gaussian then there are closed form solutions to this conditioning.  We can then predict the output ($y_*$) at an input location ($x_*$) that is not in our training data set $\mathcal{D}$.
\begin{equation}
p(y_*|\mathcal{D},x_*) = \int p(y_*|\mathbf{x},x_*,f)p(f|\mathcal{D})df
\end{equation}
The solution of this integral requires the parameters ($\theta$) of our chosen covariance function.  However, these are almost certainly unknown and must be chosen by fitting to our available data.  We maximize the marginal likelihood using a nonlinear conjugate gradient descent:
\begin{equation}
log \mbox{ } p(\mathbf{y}|\mathbf{x},\theta)
\end{equation}

Naive implementation of this has complexity $\mathcal{O}(n^3)$ for learning and inference and $\mathcal{O}(n^2)$ for prediction.  This complexity comes from computing inverses on the covariance matrix.  This has traditionally limited GP approaches to a few thousand data points.  Even in the data-poor Arctic environment, coastal data sets can be constructed with a fine spatial discretization (meters) and course temporal discretization (years) that preclude the use of brute force Gaussian process regression.

To reduce the complexity of both learning and inference, we make use of grids, Kronecker products, and circulant Toeplitz embeddings.  These techniques yield improvements of $\mathcal{O}(PN^{\frac{P+1}{P}} )$ computations and $\mathcal{O}(PN^{2P} )$ storage, for $N$ datapoints and $P$ input dimensions for hyperparameter fitting \cite{Flaxman2015}.  Together with Toeplitz block structure and kernel interpolation the complexity is further reduced to $\mathcal{O}(n)$ for learning and $\mathcal{O}(n)$ prediction \cite{Wilson2015}.

In addition to the computational complexity, another challenge facing Gaussian processes is the design or choice of a suitable covariance or kernel function.  The choice of covariance function greatly determines the dynamics that the Gaussian process can capture, as it determines the probabilities of how the output can change within the input space.  These are also the same dynamics we wish to learn from the data.  We thus need a suitable choice of covariance structure that can learn the dynamics without specifying them beforehand.  Significant work in the machine learning community has gone into developing such expressive covariance structures \cite{Wilson2013,Damianou2013,Salakhutdinov2008}.  In order to maximize the dynamics that can be learned from past data we make use of spectral mixture kernels \cite{Wilson2013} in the temporal dimension combined with several more standard kernel choices for additional structure. The spectral mixture kernel is defined as:
\begin{equation}
k(x-x') = \sum_{q=1}^Q w_q e^{-2\pi^2(x-x')^2v_q}\cos(2\pi(x-x')\mu_q).
\end{equation}
This choice of structure is equivalent to fitting the Fourier transform of the covariance with a limited number of sinusoid terms.  The hyper parameters that require fitting are thus the frequency ($\mu_i$), weights ($w_i$), and lengthscale ($v_i$) of the $Q$ sinusoids.  If provided with an initial estimate, the hyperparameters can be fit using standard gradient descent algorithms.

\subsection{GP Learning}
To calibrate our models, we partition the dataset into fit and prediction sets.  Models are trained using data from all years up to and including 2012. %[shouldn't this be 2012?].  
Models are validated by comparing the predicted 2016 coastline to the measured coastline.  Models for future scenarios are trained using all data including 2016. 

For each model calibration, the GP must learn the hyperparameters of the chosen covariance functions. The choice of covariance function to fit is a significant design choice that must consider both the spatial and temporal dimensions.  For the spatial dimension, the selected covariance function consists of the sum of a rational quadratic with an automatic relevance distance function and a squared exponential with an automatic distance function.  The rational quadratic is added to capture effects from other nearby transects (such as cross-shore erosion at one transect leading to deposition at a neighboring transect).  The squared exponential enforces a smoothness preference for neighboring coastline transects.

For the temporal dimension the chosen covariance function is the sum of a spectral mixture kernel with $8$ components, a white noise term, and a non-stationary linear covariance.  By including the non-stationary linear covariance function, we are able to capture a constantly changing coastal position with respect to time, which makes future predictions much more plausible.  The spectral mixture kernel provides a mechanism to learn unspecified dynamics with respect to time.  We include this term in order to capture many of the unmeasured variables that influence erosion rates.

For the models where near-shore temperature data is included as a third dimension, the covariance function consists of the sum of a rational quadratic with an automatic relevance distance function and a squared exponential with an automatic distance function (identical to the spatial dimension).  We thus enforce smoothness along the temperature input dimension.  

We used a Gaussian likelihood function and a constant mean function.  The likelihood function gives the probability of the observations given the parameters and the mean function acts as a non-zero, stationary offset. For hyperparameter learning we use nonlinear conjugate gradient descent for 1000 iterations with a convergence tolerance of $1 \times 10^{-3}$.  All computations are carried out using Matlab and the Gaussian Process Machine Learning Toolbox \cite{Rasmussen:2010:GPM:1756006.1953029}.

\subsection{Comparison Models}
To evaluate the performance of the GP models, we compare the results of the Gaussian process approach to a linear model fit to each transect as well as a nonlinear model fit to each transect (where sufficient data are available).  To include temperature effects we also fit a nonlinear model using near shore temperature.  For most of the Arctic coastline, linear interpolation provides the only source of prediction available \cite{Gibbs2015}.

The linear model is fit to each transect individually and is of the form:
\begin{equation}
\mbox{distance from baseline} = a(\mbox{year})+b 
\end{equation}
The nonlinear model is also fit to each transect individually and is of the form:
\begin{equation}
\mbox{distance from baseline} = a(\mbox{year})^b+c 
\end{equation}
The nonlinear temperature model is fit to each transect and is of the form:
\begin{equation}
\mbox{distance from baseline} = a(\mbox{year})+b(\mbox{temp})^c+d 
\end{equation}
The nonlinear model has three parameters which must be fit, and also requires an initial guess in order to iterate to an optimized value.  It thus requires more measurements which limits its applicability.  If fewer than three measurements were available to fit the nonlinear model at a given transect, the linear model was used for that transect.  The same approach is used with the nonlinear temperature model, which requires four measurements to fit, if fewer measurements are available the linear model is used at that transect.

\section{Results}
In this section, we present results for the Gaussian process models compared against the linear and nonlinear regression models.  Both GP models use the temporal and spatial dimensions of the measured coastline data, but one of them also includes the historical simulated temperature data. We evaluated each model by calculating the prediction error, root mean square error (RMSE), at every transect and for the error vector as a whole. Since errors can vary substantially over space, we present transect-level errors graphically and include the overall RMSE as a way to quickly evaluate the overall predictive power of each model.

The RMSE values for all models over all coastlines are shown in Table \ref{tab:predRMSE}.  The Gaussian process method shows a significant decrease in RMSE over all analyzed coastlines.  Explicitly including the spatial dependence of transect locations improves the predictive power of the GP models compared to the transect-specific linear and nonlinear regression models. The Gaussian process is also able to better capture the nonlinear and spatially varying change in erosion rates over time.  

\begin{figure}[!t]
\centering
%\noindent
  \resizebox{8cm}{!}{% This file is generated by the MATLAB m-file laprint.m. It can be included
% into LaTeX documents using the packages graphicx, color and psfrag.
% It is accompanied by a postscript file. A sample LaTeX file is:
%    \documentclass{article}\usepackage{graphicx,color,psfrag}
%    \begin{document}\input{Spatial2016}\end{document}
% See http://www.mathworks.de/matlabcentral/fileexchange/loadFile.do?objectId=4638
% for recent versions of laprint.m.
%
% created by:           LaPrint version 3.16 (13.9.2004)
% created on:           31-Aug-2017 16:31:21
% eps bounding box:     15 cm x 10.1877 cm
% comment:              
%
\begin{psfrags}%
\psfragscanon%
%
% text strings:
\psfrag{s07}[t][t]{\color[rgb]{0.15,0.15,0.15}\setlength{\tabcolsep}{0pt}\begin{tabular}{c}UTM X (meters) $\times 10^{5}$\end{tabular}}%
\psfrag{s08}[b][b]{\color[rgb]{0.15,0.15,0.15}\setlength{\tabcolsep}{0pt}\begin{tabular}{c}UTM Y (meters) $\times 10^{6}$\end{tabular}}%
%
% axes ticklabel color:
\color[rgb]{0.15,0.15,0.15}%
%
% xticklabels:
\psfrag{x01}[t][t]{4.738}%
\psfrag{x02}[t][t]{4.74}%
\psfrag{x03}[t][t]{4.742}%
\psfrag{x04}[t][t]{4.744}%
\psfrag{x05}[t][t]{4.746}%
\psfrag{x06}[t][t]{4.748}%
\psfrag{x07}[t][t]{}%
\psfrag{x08}[t][t]{4.67}%
\psfrag{x09}[t][t]{4.68}%
\psfrag{x10}[t][t]{4.69}%
\psfrag{x11}[t][t]{4.7}%
\psfrag{x12}[t][t]{4.71}%
\psfrag{x13}[t][t]{4.72}%
\psfrag{x14}[t][t]{4.73}%
\psfrag{x15}[t][t]{4.74}%
\psfrag{x16}[t][t]{}%
%
% yticklabels:
\psfrag{v01}[r][r]{7.8644}%
\psfrag{v02}[r][r]{7.8646}%
\psfrag{v03}[r][r]{7.8648}%
\psfrag{v04}[r][r]{7.865}%
\psfrag{v05}[r][r]{7.8652}%
\psfrag{ypower1}[Bl][Bl]{}%
\psfrag{v06}[r][r]{7.864}%
\psfrag{v07}[r][r]{7.865}%
\psfrag{v08}[r][r]{7.866}%
\psfrag{v09}[r][r]{7.867}%
\psfrag{v10}[r][r]{7.868}%
\psfrag{v11}[r][r]{7.869}%
\psfrag{v12}[r][r]{7.87}%
\psfrag{ypower2}[Bl][Bl]{}%
%
% Figure:
\resizebox{12cm}{!}{\includegraphics{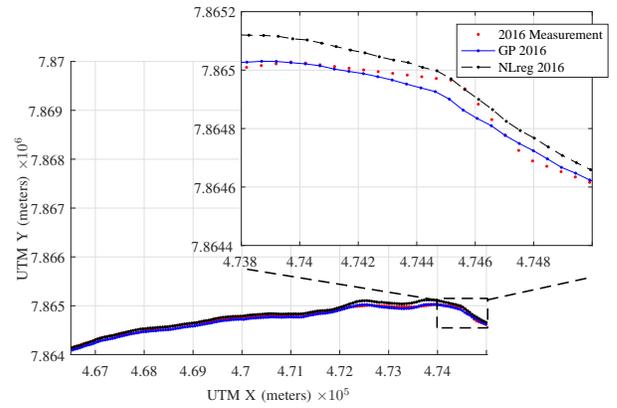}}%
\end{psfrags}%
%
% End Spatial2016.tex
}
\caption{Temporal Spatial model validation for 2016 and viewable inset.}
\label{fig:spatial2016}
\end{figure}

Figure \ref{fig:spatial2016} shows the west section of coastline and zoomed-in subsegment for the linear, nonlinear, and spatiotemporal Gaussian process models.  Linear and nonlinear regression techniques consistently under-estimate erosion rates at the majority of the transects, while the Gaussian process is prone to over-estimating erosion rates, though overall it predicts closer to the actual measurement.
\begin{table}[h]
\begin{center}
\caption{RMSE Comparison for 2016 Prediction} \label{tab:predRMSE}
\begin{tabular}{lcc}
\hline
\cellcolor[gray]{0.8}Method & \multicolumn{2}{c}{\cellcolor[gray]{0.8}RMSE}  \\ \cline{2-3} & West Section & East Section   \\  
%\multicolumn{3}{l}{\cellcolor[gray]{0.8}GP Spatial Temporal Model} \\
\hline
 Linear Regression & $53.99$ & $49.41$    \\
 Nonlinear Regression & $50.34$ & $160.17$   \\
 Nonlinear Regression Temperature & $33.22$ & $96.52$   \\
 GP Temporal Spatial & $35.17$ & $32.62$     \\
 GP Temporal Spatial Temperature & $23.79$ & $68.55$  \\
 \hline
\end{tabular}
\end{center}
\end{table}

The number of coastline measurements along each transect has a clear effect on the higher order models.  The eastern section has only four annual measurements along each transect.  With so few measurements the nonlinear regression methods have very high errors when compared to straight linear regression using only time as an input. The GP temperature model also fails to achieve an improvement over linear regression when so few coastlines are available to learn from.  However, even with so few measurements, the GP model using only spatio-temporal inputs still shows a significant improvement over simple linear regression.

\begin{figure}[h]
\centering
%\noindent
  \resizebox{8cm}{!}{% This file is generated by the MATLAB m-file laprint.m. It can be included
% into LaTeX documents using the packages graphicx, color and psfrag.
% It is accompanied by a postscript file. A sample LaTeX file is:
%    \documentclass{article}\usepackage{graphicx,color,psfrag}
%    \begin{document}\input{FutureComp}\end{document}
% See http://www.mathworks.de/matlabcentral/fileexchange/loadFile.do?objectId=4638
% for recent versions of laprint.m.
%
% created by:           LaPrint version 3.16 (13.9.2004)
% created on:           31-Aug-2017 16:41:46
% eps bounding box:     15 cm x 10.0933 cm
% comment:              
%
\begin{psfrags}%
\psfragscanon%
%
% text strings:
\psfrag{s02}[b][b]{\color[rgb]{0.15,0.15,0.15}\setlength{\tabcolsep}{0pt}\begin{tabular}{c}UTM Y (meters)\end{tabular}}%
\psfrag{s03}[t][t]{\color[rgb]{0.15,0.15,0.15}\setlength{\tabcolsep}{0pt}\begin{tabular}{c}UTM X (meters)\end{tabular}}%
%
% axes ticklabel color:
\color[rgb]{0.15,0.15,0.15}%
%
% xticklabels:
\psfrag{x01}[t][t]{4.738}%
\psfrag{x02}[t][t]{4.74}%
\psfrag{x03}[t][t]{4.742}%
\psfrag{x04}[t][t]{4.744}%
\psfrag{x05}[t][t]{4.746}%
\psfrag{x06}[t][t]{4.748}%
\psfrag{x07}[t][t]{\shortstack{4.75\\$\times 10^{5}\ $}}%
%
% yticklabels:
\psfrag{v01}[r][r]{7.8644}%
\psfrag{v02}[r][r]{7.8645}%
\psfrag{v03}[r][r]{7.8646}%
\psfrag{v04}[r][r]{7.8647}%
\psfrag{v05}[r][r]{7.8648}%
\psfrag{v06}[r][r]{7.8649}%
\psfrag{v07}[r][r]{7.865}%
\psfrag{v08}[r][r]{7.8651}%
\psfrag{ypower1}[Bl][Bl]{$\times 10^{6}$}%
%
% Figure:
\resizebox{12cm}{!}{\includegraphics{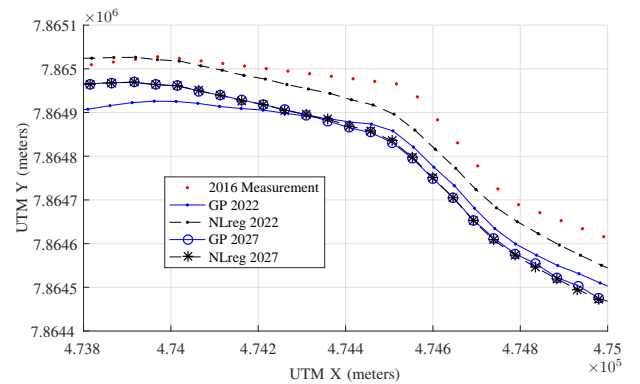}}%
\end{psfrags}%
%
% End FutureComp.tex
}
\caption{Temporal Spatial model predictions for 2022 and 2027.}
\label{fig:future_comp}
\end{figure}

We are also able to generate future erosion rates with the hope that these forecasts can be validated in the future. The forecast goal of the algorithm is annual to decadal, so we make coastline forecasts for 2022 and 2027, 5 and 10 years into the future respectively.  The results of these forecasts are shown for the zoomed-in subsegment coastline (Figure \ref{fig:future_comp}).  This is not a scenario-based forecast as no environmental data are included.  Instead, the model simply assumes that historical erosion processes will continue into the future.  The nonlinear and linear (not plotted) regression both show  under estimation of erosion. For the five-year forecast (2022), both the nonlinear regression and linear regression models forecast a coastline that is behind the 2016 measurement.
\vspace{.5cm}

The Gaussian process forecast for 2022 shows continued erosion past the 2016 measurement.  However when the forecast time frame is extended to 2027, the Gaussian process method forecasts both erosion and deposition such that the resulting coastline is nearly identical to the nonlinear regression model.  This is due to stationarity embedded within the covariance function.  As the temporal dimension moves farther beyond the training data, it will resort to the only non-stationary term provided which is strictly linear% [but we are comparing to the nonlinear model -- how to reconcile this issue?]
. This diminishes the contribution of the learned effects until the only remaining effect is a linear one.  As we forecast farther outside the fit window, the Gaussian process method effectively converges to the linear prediction (which are visually indistinguishable from the nonlinear regression results shown in Figure \ref{fig:future_comp}).

\begin{figure}[!t]
\centering
%\noindent
  \resizebox{8cm}{!}{% This file is generated by the MATLAB m-file laprint.m. It can be included
% into LaTeX documents using the packages graphicx, color and psfrag.
% It is accompanied by a postscript file. A sample LaTeX file is:
%    \documentclass{article}\usepackage{graphicx,color,psfrag}
%    \begin{document}\input{SpatialTempNS}\end{document}
% See http://www.mathworks.de/matlabcentral/fileexchange/loadFile.do?objectId=4638
% for recent versions of laprint.m.
%
% created by:           LaPrint version 3.16 (13.9.2004)
% created on:           31-Aug-2017 17:19:16
% eps bounding box:     15 cm x 10.2208 cm
% comment:              
%
\begin{psfrags}%
\psfragscanon%
%
% text strings:
\psfrag{s03}[t][t]{\color[rgb]{0.15,0.15,0.15}\setlength{\tabcolsep}{0pt}\begin{tabular}{c}UTM X (meters $\times 10^{5}$)\end{tabular}}%
\psfrag{s04}[b][b]{\color[rgb]{0.15,0.15,0.15}\setlength{\tabcolsep}{0pt}\begin{tabular}{c}UTM Y (meters $\times 10^{6}$)\end{tabular}}%
%
% axes ticklabel color:
\color[rgb]{0.15,0.15,0.15}%
%
% xticklabels:
\psfrag{x01}[t][t]{4.738}%
\psfrag{x02}[t][t]{4.74}%
\psfrag{x03}[t][t]{4.742}%
\psfrag{x04}[t][t]{4.744}%
\psfrag{x05}[t][t]{4.746}%
\psfrag{x06}[t][t]{4.748}%
\psfrag{x07}[t][t]{}%
\psfrag{x08}[t][t]{4.67}%
\psfrag{x09}[t][t]{4.68}%
\psfrag{x10}[t][t]{4.69}%
\psfrag{x11}[t][t]{4.7}%
\psfrag{x12}[t][t]{4.71}%
\psfrag{x13}[t][t]{4.72}%
\psfrag{x14}[t][t]{4.73}%
\psfrag{x15}[t][t]{4.74}%
\psfrag{x16}[t][t]{}%
%
% yticklabels:
\psfrag{v01}[r][r]{7.8644}%
\psfrag{v02}[r][r]{7.8646}%
\psfrag{v03}[r][r]{7.8648}%
\psfrag{v04}[r][r]{7.865}%
\psfrag{v05}[r][r]{7.8652}%
\psfrag{ypower1}[Bl][Bl]{}%
\psfrag{v06}[r][r]{7.864}%
\psfrag{v07}[r][r]{7.865}%
\psfrag{v08}[r][r]{7.866}%
\psfrag{v09}[r][r]{7.867}%
\psfrag{v10}[r][r]{7.868}%
\psfrag{v11}[r][r]{7.869}%
\psfrag{v12}[r][r]{7.87}%
\psfrag{ypower2}[Bl][Bl]{}%
%
% Figure:
\resizebox{12cm}{!}{\includegraphics{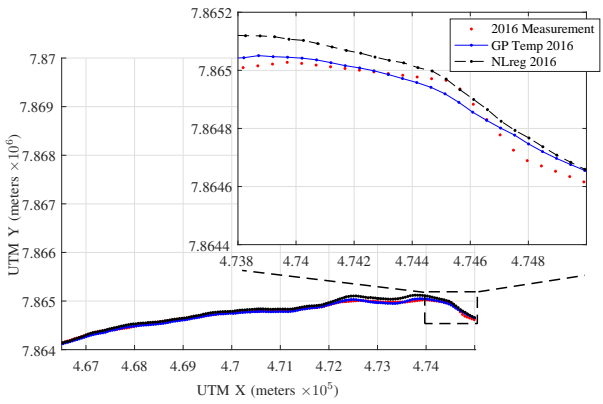}}%
\end{psfrags}%
%
% End SpatialTempNS.tex
}
\caption{Temporal Spatial Temperature model validation for 2016 and viewable inset.}
\label{fig:spatialNS2016}
\end{figure}

This convergence can be eliminated by including environmental data into the forecasts.  By including near-shore temperature data, we are able to further reduce our model error to less than half the total error present in the nonlinear regression model (Table \ref{tab:predRMSE})% [this is not really a fair comparison though, since the nonlinear model does not have an added temperature predictor!]
.  For our coastline subsegment in our visualizations, the addition of temperature improves the fit for some transects, but not all (Figure \ref{fig:spatialNS2016}).

\begin{figure}[!t]
\centering
%\noindent
  \resizebox{8cm}{!}{% This file is generated by the MATLAB m-file laprint.m. It can be included
% into LaTeX documents using the packages graphicx, color and psfrag.
% It is accompanied by a postscript file. A sample LaTeX file is:
%    \documentclass{article}\usepackage{graphicx,color,psfrag}
%    \begin{document}\input{STNSFutureOptim}\end{document}
% See http://www.mathworks.de/matlabcentral/fileexchange/loadFile.do?objectId=4638
% for recent versions of laprint.m.
%
% created by:           LaPrint version 3.16 (13.9.2004)
% created on:           31-Aug-2017 17:33:49
% eps bounding box:     15 cm x 9.4464 cm
% comment:              
%
\begin{psfrags}%
\psfragscanon%
%
% text strings:
\psfrag{s02}[t][t]{\color[rgb]{0.15,0.15,0.15}\setlength{\tabcolsep}{0pt}\begin{tabular}{c}UTM X (meters)\end{tabular}}%
\psfrag{s03}[b][b]{\color[rgb]{0.15,0.15,0.15}\setlength{\tabcolsep}{0pt}\begin{tabular}{c}UTM Y (meters)\end{tabular}}%
%
% axes ticklabel color:
\color[rgb]{0.15,0.15,0.15}%
%
% xticklabels:
\psfrag{x01}[t][t]{4.738}%
\psfrag{x02}[t][t]{4.74}%
\psfrag{x03}[t][t]{4.742}%
\psfrag{x04}[t][t]{4.744}%
\psfrag{x05}[t][t]{4.746}%
\psfrag{x06}[t][t]{4.748}%
\psfrag{x07}[t][t]{\shortstack{4.75\\$\times 10^{5}\ $}}%
%
% yticklabels:
\psfrag{v01}[r][r]{7.8644}%
\psfrag{v02}[r][r]{7.8645}%
\psfrag{v03}[r][r]{7.8646}%
\psfrag{v04}[r][r]{7.8647}%
\psfrag{v05}[r][r]{7.8648}%
\psfrag{v06}[r][r]{7.8649}%
\psfrag{v07}[r][r]{7.865}%
\psfrag{v08}[r][r]{7.8651}%
\psfrag{ypower1}[Bl][Bl]{$\times 10^{6}$}%
%
% Figure:
\resizebox{12cm}{!}{\includegraphics{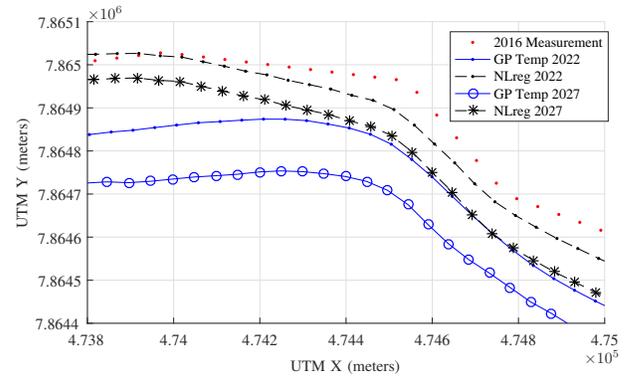}}%
\end{psfrags}%
%
% End STNSFutureOptim.tex
}
\caption{Temporal Spatial Temperature model predictions for 2022 and 2027, optimistic scenario (2.6RCP).}
\label{fig:future_comp_optim}
\end{figure}

A more noticeable benefits of including temperature data is to enable scenario-based coastline forecasts that are more believable, i.e. they do not revert to a simplistic linear model. This is especially important when generating forecasts at the decadal time frame.  To generate these future scenarios, we use downsampled forecasted near-shore temperature data.  For even the optimistic scenario (2.6RCP) the Gaussian process predicts higher erosion rates than the simpler spatio-temporal model. Figure \ref{fig:future_comp_optim} shows the forecast coastlines for this future scenario characterized by low CO$_2$ emissions.

\begin{figure}[!b]
\centering
%\noindent
  \resizebox{8cm}{!}{% This file is generated by the MATLAB m-file laprint.m. It can be included
% into LaTeX documents using the packages graphicx, color and psfrag.
% It is accompanied by a postscript file. A sample LaTeX file is:
%    \documentclass{article}\usepackage{graphicx,color,psfrag}
%    \begin{document}\input{STNSFuturePess}\end{document}
% See http://www.mathworks.de/matlabcentral/fileexchange/loadFile.do?objectId=4638
% for recent versions of laprint.m.
%
% created by:           LaPrint version 3.16 (13.9.2004)
% created on:           31-Aug-2017 18:05:09
% eps bounding box:     15 cm x 9.1941 cm
% comment:              
%
\begin{psfrags}%
\psfragscanon%
%
% text strings:
\psfrag{s03}[t][t]{\color[rgb]{0.15,0.15,0.15}\setlength{\tabcolsep}{0pt}\begin{tabular}{c}UTM X (meters)\end{tabular}}%
\psfrag{s04}[b][b]{\color[rgb]{0.15,0.15,0.15}\setlength{\tabcolsep}{0pt}\begin{tabular}{c}UTM Y (meters)\end{tabular}}%
%
% axes ticklabel color:
\color[rgb]{0.15,0.15,0.15}%
%
% xticklabels:
\psfrag{x01}[t][t]{4.738}%
\psfrag{x02}[t][t]{4.74}%
\psfrag{x03}[t][t]{4.742}%
\psfrag{x04}[t][t]{4.744}%
\psfrag{x05}[t][t]{4.746}%
\psfrag{x06}[t][t]{4.748}%
\psfrag{x07}[t][t]{\shortstack{4.75\\$\times 10^{5}\ $}}%
%
% yticklabels:
\psfrag{v01}[r][r]{7.8644}%
\psfrag{v02}[r][r]{7.8645}%
\psfrag{v03}[r][r]{7.8646}%
\psfrag{v04}[r][r]{7.8647}%
\psfrag{v05}[r][r]{7.8648}%
\psfrag{v06}[r][r]{7.8649}%
\psfrag{v07}[r][r]{7.865}%
\psfrag{v08}[r][r]{7.8651}%
\psfrag{ypower1}[Bl][Bl]{$\times 10^{6}$}%
%
% Figure:
\resizebox{12cm}{!}{\includegraphics{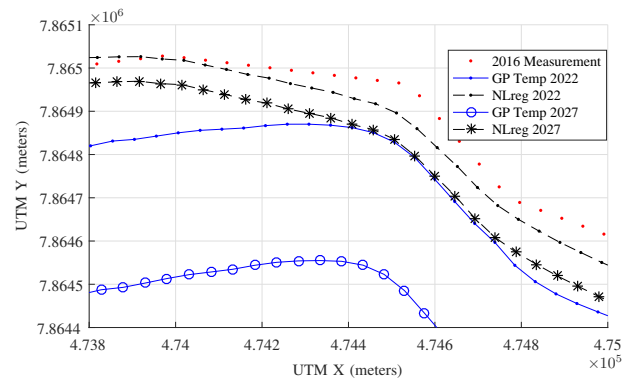}}%
\end{psfrags}%
%
% End STNSFuturePess.tex
}
\caption{Temporal Spatial Temperature model predictions for 2022 and 2027, pessimistic scenario (8.5RCP).}
\label{fig:future_comp_pess}
\end{figure}

When temperatures increase according to a more pessimistic future scenario (8.5RCP), the model forecasts higher erosion rates.  This increase is shown on the viewable subsegment in Figure \ref{fig:future_comp_pess}.  Given the superior performance of this spatio-temporal GP model with a temperature input, these future forecasts of substantial coastal erosion are likely to be more accurate and should be considered as plausible outcomes for future scenarios by Arctic communities and stakeholders.

\section{Conclusion}
In this work, we demonstrate the benefits of applying machine learning to study long-term (decadal) coastal morphology over coastlines exhibiting variable erosion mechanisms.  The Gaussian process model accuracies were considerably better than competing linear and nonlinear regression models. Overall, model accuracy was best in higher-density areas with more frequent observations in the training set. %[is it worth quantifying this statement by generating RMSEs for different portions of the coast?]
%[yes, I think we need to quantify this statement or remove it]
The use of a Gaussian process provides a rapid and effective way to include additional measurements as they become available to generate future predictions on an annual timescale.  Data for much of the Arctic coastline is sparse, making erosion forecasts difficult for these regions, no matter the choice of modeling framework.

As additional satellite imagery becomes accessible and more frequent coastlines can be extracted, the coastal areas that can be modeled with confidence will increase.  Many of the historical coastlines for this work relied on manual photo interpretation, but with high resolution imagery, the modeling approach can be extended to larger coastlines by relying on automatic coastal detection algorithms \cite{Alesheikh2007}.  The potential for rapidly incorporating new data and predicting very long coastlines hundreds to thousands of kilometers in length is a particular strength of the machine learning approach.  Although we did not experience any computational constraints for our study area data set, as data availability for the Arctic increases and longer coastlines can be modeled, applying Gaussian processes will become more challenging.  One way to mitigate these expected computational limits is to reduce the coastline sampling frequency, with care given to retain spatial dependence between neighboring transects since this information is important for the machine learning methodology.

We also showed that by including hindcasted near-shore temperature data, model accuracy can be improved substantially.  Furthermore, inclusion of such environmental data enables generating forecasts in line with future plausible scenarios, providing an additional tool for communities and decision makers.

For future work, we hope to include other hindcasted variables that are known to affect erosion rates, such as sea ice extent, coastal particle size, and aggregate wind and wave measurements.% [the obvious critique here is why have we not already done this...]
 Some of these measures are very difficult to obtain for the Arctic, and require the imposition of additional assumptions to obtain fine-resolution estimates.  We plan to explore the predictive power of these additional covariates to assess their relative importance and to gain insight into the underlying physical processes that govern Artic coastal erosion.  %Including these additional regressors will also allow for scenario-based forecasts.
% or
%\appendices
%% you can choose not to have a title for an appendix
%% if you want by leaving the argument blank
%\section{}
\section*{Acknowledgment}
Special thanks to Katherine Barnhart for generously providing the GPS track data for the Drew Point coastline.  Thanks to Ben Jones for providing coastline shapefiles and thanks to Jinlun Zhang for providing temperature data.  This work was funded, in part, by the University of Alaska Anchorage ConocoPhillips Arctic Science and Engineering Award.
\ifCLASSOPTIONcaptionsoff
  \newpage
\fi


%\bibliographystyle{IEEEtran}
\bibliographystyle{IEEEtran} 
\bibliography{coastfrank}


%\end{IEEEbiography}
%% insert where needed to balance the two columns on the last page with
%%\newpage


%\vfill


\end{document}